\documentclass[conference]{IEEEtran}
\IEEEoverridecommandlockouts
\usepackage{cite}
\usepackage{amsmath,amssymb,amsfonts}
\usepackage{algorithmic}
\usepackage{graphicx}
\usepackage{textcomp}
\usepackage{xcolor}
\usepackage{url}
\usepackage{multirow}

\def\BibTeX{{\rm B\kern-.05em{\sc i\kern-.025em b}\kern-.08em
    T\kern-.1667em\lower.7ex\hbox{E}\kern-.125emX}}
\begin{document}

\title{Exploring LLM Support for Generating IEC 61131-3 Graphic Language Programs
}

\author{\IEEEauthorblockN{Yimin Zhang}
\IEEEauthorblockA{\textit{CISTER / Faculty of Engineering} \\
\textit{University of Porto}\\
Porto, Portugal \\
0009-0005-0746-315X}
\and
\IEEEauthorblockN{Mario de Sousa}
\IEEEauthorblockA{\textit{Faculty of Engineering} \\
\textit{University of Porto}\\
Porto, Portugal \\
0000-0001-7200-1705}
}

\maketitle

\begin{abstract}
The capabilities demonstrated by Large Language Models (LLMs) inspire researchers to integrate them into industrial production and automation.
In the field of Programmable Logic Controller (PLC) programming, previous researchers have focused on using LLMs to generate Structured Text (ST) language, and created automatic programming workflows based on it. The IEC 61131 graphic programming languages, which still has the most users~\cite{Technavio}, have however been overlooked.

In this paper we explore using LLMs to generate graphic languages in ASCII art to provide assistance to engineers. Our series of experiments indicate that, contrary to what researchers usually think, it is possible to generate a correct Sequential Function Chart (SFC) for simple requirements when LLM is provided with several examples.
On the other hand, generating a Ladder Diagram (LD) automatically remains a challenge even for very simple use cases.
The automatic conversion between LD and SFC without extra information also fails when using prompt engineering alone.
\end{abstract}


\section{Introduction} \label{sec:introduction}
The rise of ChatGPT and other Large Language Models (LLMs) is changing people's work and lifestyle. The same is true in industry.
More and more people began to discuss what role LLMs can play in industrial domain.
The most optimistic estimates even claim that a breakthrough will be achieved within a few years.

\subsection{LLMs and Industry}
LLMs can liberate engineers from heavy paperwork, allowing them to focus on their work.
A fine-tuned model within an enterprise may replace some after-sales engineers, undertaking preliminary troubleshooting tasks.
In the field of Artificial Intelligence (AI), these systems are often mentioned as ``agents''.
LLM is on par with humans in general purpose programming languages~\cite{achiam2023gpt}, which has made some engineers worried about unemployment.
Several vendors have already introduced support for AI in the context of industrial automation (Siemens~\cite{SiemensAI}, CODESYS~\cite{CODESYSAI}, Beckhoff~\cite{BeckhoffAI}, etc.).

Unlike the daily use of LLMs, industrial applications face hard requirements and place more emphasis on precision, clear instructions and reliability.
Every procedure and production process needs to be carefully designed to avoid damage.
While we believe that LLMs can provide domain knowledge, we cannot fully rely on them.
In other words, a LLM based tool is best used as a helpful assistant.
Their outputs need to be rigorously validated.

\subsection{LLMs and PLC programming}
Three of the five programming languages defined in IEC 61131-3 are graphic languages - Ladder Diagram (LD), Sequential Function Chart (SFC) and Function Block Diagram (FBD).
LD dominates among the 5 languages, accounting for more than 80\% of the global use~\cite{Technavio}.
Its close resemblance to electrical circuits makes LD popular.
Technicians who are familiar with electrical circuits but lack a knowledge of general-purpose computer languages, such as C, Java, or Python, use these graphic languages to develop PLC software.

However, LLM is typically considered a language model, which means that it's better to process textual information.
The graphical nature of LD, SFC, and FBD limit the application of LLM in PLC programming.
Recent researchers (\cite{AbderrahmanePreliminary},~\cite{koziolek2023chatgpt},~\cite{fakih2024llm4plc}) have therefore focused on Structured Text (ST), while overlooking the most commonly used LD, SFC, and FBD (instruction List (IL) is rarely used in practice.)

However, the standard does specify these graphic languages in terms of ASCII art~\cite{iec61131-3v3}.
In actual industrial use IEC 61131-3 editors offer GUIs (Graphical User Interfaces) that rely on graphically similar elements.
Nevertheless, the most important is the logic behind the graphic languages rather than the implementation form.
In this paper we explore the assistance that LLMs can provide in PLC graphical programming starting from the ASCII art.
In so doing we found that LLM can generate semantically and syntactically correct SFCs for simple control logic.

The rest of the paper is structured as follows.
In Section~\ref{sec:SOTA}, we review the related work. We describe the methodology in Section~\ref{sec:methodology}. Section~\ref{sec:experiment} showcases the experiments we did, and Section~\ref{sec:conclusion} concludes the paper.

\section{State of the Art} \label{sec:SOTA}

Academic or industrial teams have released many LLMs for code generation, e.g., GitHub Copilot~\cite{GitHubCopilot}, CodeT~\cite{chen2022codet}, Gemini~\cite{geminiteam2023gemini}, Code Llama~\cite{rozière2024code}. Some of them have achieved good performance in relevant tests.
However, these LLMs usually target towards general-purpose programming languages, such as Python and Java, and rarely involve PLC programming.

The team from ABB~\cite{koziolek2023chatgpt} systematically examines GPT-4's capability to generate ST code. They build up a preliminary benchmark containing 10 possible scenarios for industrial applications that could be used for further LLM comparison. Each category contains ten different problems.
To evaluate the quality of GPT-4's answers, they designed a scoring system, measuring prompt difficulties, syntax correctness, etc.
They summarize and collect the results in a workbook, providing a preliminary benchmark.

Through their experiments, the team reached some interesting conclusions.
We will cite several key conclusions that are important to this study as follows:
\begin{itemize}
    \item ChatGPT can generate syntactically correct ST;
    \item ChatGPT can provide domain knowledge;
    \item Prompt fidelity largely determines answer fidelity;
\end{itemize}

However, this research has some limitations:
\begin{itemize}
    \item Lack of basic strategies to get better results such as providing system / assistance messages.
    \item Most of the prompts are zero-shot learning, without attempting few-shot learning.
\end{itemize}

In~\cite{AbderrahmanePreliminary}, Abderrahmane et al. propose a workflow, using LLM, Reinforcement Learning (RL), Natural Language Processing (NLP), etc., to automate PLC programming in Structured Text.
However, they did not provide sufficient experiments and data, leaving such a pipeline still in the conceptual stage.

On the contrary, in~\cite{fakih2024llm4plc}, through a series of experiments, Fakih et al. demonstrated an LLM-augmented workflow that automates PLC programming also in Structured Text.
The workflow includes syntax checking, formal verification, and human supervision to improve the accuracy and usability of automatically generated code.
They test the pipeline on GPT-3.5, GPT-4, Code Llama-7B, etc., finally achieving an improvement of generation success rate from 47\% to 72\%.
However, they test on a standard library, raising the concerns of the model memorizing the standard library corpus.

\section{Methodology} \label{sec:methodology}

The purpose of our experiments is to explore LLMs' capabilities in generating PLC graphic programming languages (LD, SFC and FBD). Due to the nature of our research interests, for the moment we have focused our experiments on situations better modeled as discrete event systems, for which FBD is not the most appropriate language.

\subsection{Prompts / Questions} \label{subsect:prompts}
The experiments conducted by~\cite{koziolek2023chatgpt} revealed that ChatGPT can generate syntactically correct IEC 61131-3 Structured Text code. A very basic test to generate SFC was not successful.
For the sake of comparison we continue to use the benchmark\footnote{\url{https://github.com/hkoziolek/control-logic-generation-prompts}} they provided in the first  experiments.
We limited our research to the Categorie~3 to~5 that are related to discrete event systems i.e. PLC Programming Exercises, Process Control and Sequential Control.
The prompts used from Experiment~3 to Experiment~6 will be discussed later in Section~\ref{exp:LD_Detail}.

\subsection{LLM Model}
Considering its popularity and ease of comparison, we choose OpenAI API, ``gpt-4-turbo-preview'', for our experimentation.
The version is ``gpt-4-0125-preview'' for which OpenAI claims that ``this model completes tasks like code generation more thoroughly than the previous preview model and is intended to reduce cases of `laziness' where the model does not complete a task~\cite{gpt-4-turbo-preview}.''


\subsection{Few-shot Learning}
Previous preliminary prompt engineering, which could be considered as zero-shot learning, has already demonstrated some surprising conclusions in PLC programming. However, LLMs are few-shot learners~\cite{brown2020language}.
It is generally believed that the performance of few-shot learning is better than zero-shot learning, which prompts our curiosity: can LLMs handle more complex PLC programming when given examples?

\subsection{Evaluation}
At present, we are not aware of any compiler that can address programs in ASCII art format, even though the IEC 61131-3 standard itself formally defines ASCII art representations for all graphical languages, including LD. It is therefore difficult to automate the evaluation of the correctness of a LLM's outputs. One approach is to manually input LLM's answers into an IDE, but this would entail a huge workload and go against the initial purpose. The method we adopt is manual judgment and scoring of the answers, which is also made possible by the limited number of discrete events in the problems being analysed.

For transparency all outputs from LLM interface are published on GitHub\footnote{\url{https://github.com/yimin-up/LLM_IEC-61131_Graphic}}.

\begin{table}[!htbp]
    \centering
    \caption{Scoring instructions}
    \begin{tabular}{|c|c|}
    \hline
        \textbf{Score} & \textbf{Meaning} \\ \hline
        \textbf{-1} & Don't provide a solution in required language. \\ \hline
        \textbf{0} & Semantically or syntactically Wrong. \\ \hline
        \textbf{0.5} & Wrong, but engineers can understand. \\ \hline
        \textbf{1} & Semantically and syntactically correct. \\ \hline
    \end{tabular}
    \label{table:scoring}
\end{table}

Regarding the scoring, the Table~\ref{table:scoring} outlines the rules for scoring.
It should be noted that in order to avoid giving abrupt judgments of incorrect or correct answers, we divide this into 4 levels.
Half points represent that, although not syntactically and semantically correct, engineers with basic knowledge of LD or SFC can understand the intention of the graphics. As mentioned in the introduction, this can also serve as programming assistance.

\section{Experiments} \label{sec:experiment}
A summary of all experiments is shown in Table \ref{tab:experiment_summary} describing strategies applied in each experiment.
\begin{table}[htbp]
    \centering
    \caption{Experiments Summary}
    \begin{tabular}{|c|c|c|c|c|}
    \hline
        \textbf{Index} & \textbf{System Message} & \textbf{Subtask} & \textbf{Few-shot} \\ \hline
        Experiment 1 & \checkmark & \checkmark & \checkmark \\ \hline
        Experiment 2 & \checkmark & \checkmark & \checkmark \\ \hline
        Experiment 3 & \checkmark & x & \checkmark \\ \hline
        Experiment 4 & \checkmark & x & \checkmark \\ \hline
        Experiment 5 & \checkmark & x & \checkmark \\ \hline
        Experiment 6 & \checkmark & x & \checkmark \\ \hline
        Experiment 7 & \checkmark & x & \checkmark \\ \hline
    \end{tabular}
    \label{tab:experiment_summary}
\end{table}

\subsection{Experiment 1: PLC programs in Ladder Diagram} \label{exp:LD}
In this experiment, we require GPT-4 model to provide the Ladder Diagram solutions for 30 questions (10 questions from each of 3 categories). We use the scoring mechanism mentioned earlier and summarize the results in Table~\ref{tab:LD}. Due to space constraints only the 10 questions of Category 3 are displayed here for ease of comparison.

We will compare the following three strategies:
\begin{itemize}
    \item System messages: \textit{You are an AI expert in PLC (Programmable Logic Controller) programming. You are familiar with the international standard IEC 61131 for PLCs. You will be provided with some design requirements. Your task is to write IEC 61131-3 Ladder Diagram programs in ASCII art according to the requirements.}
    \item System messages divided into steps: \textit{Follow these steps to fulfil the requirements:
        \begin{itemize}
        \item Step 1: Define and list the inputs and outputs of the Ladder Diagram, and the standard symbols they use.
        \item Step 2: Give the Ladder Diagram program.
        \item Step 3: Give the logical expressions for each rung.
        \item Step 4: Briefly describe the state machine of the Ladder Diagram.
        \item Step 5: Check the LD you generate according to the IEC 61131-3 standard.
        \end{itemize}}
    \item Few-shot learning: provide 3 examples in prompts.
\end{itemize}

The 3 examples are:\\
\textbf{Example 1} \label{exp:LD01}\\
User messages: \textit{Use start / stop button to control a motor. When the start button is pressed, the motor starts. When the stop button is pressed, the motor stops.}\\
Assistant messages:
\begin{verbatim}
Below is a simple representation of
the ladder diagram in ASCII art.
```
(Start of Ladder Diagram)
|                                   |
| StartButton   StopButton  Motor   |
+----| |------+-----|/|-------( )---+
|             |                     |
|   Motor     |                     |
+----| |------+                     |
(End of Ladder Diagram)
```
\end{verbatim}
\textbf{Example 2} \label{exp:LD02}\\
User messages: \textit{Design a counter with reset control input. If reset is True, counter value is reset to the value defined by global configuration constant ResetCounterValue. If reset is False, counter is incremented every PLC cycle.}\\
Assistant messages:
\begin{verbatim}
The Ladder Diagram program in ASCII art:
```
(Start of Ladder Diagram)
|                   +---------+      |
|                   |   SEL   |      |
|    Reset          |         |      |
+-----| |-----------|G     OUT|--cnt |
|                   |         |      |
|      +-------+    |         |      |
|      |  ADD  |    |         |      |
|   1--|IN1 OUT|----|IN0      |      |
| cnt--|IN2    |    |         |      |
|      +-------+ +--|IN1      |      |
|                |  +---------+      |
|    ResetValue--+                   |
(End of Ladder Diagram)
```
\end{verbatim}
\textbf{Example 3} \label{exp:LD03}\\
User messages: \textit{Create a program for 8-bit running LED lights.}\\
Assistant messages:
\begin{verbatim}
Ladder Diagram Program:
```
(Start of Ladder Diagram)
|   start                     status |
+----| |------------------------(S)--+
|                                    |
|    stop                     status |
+----| |------------------------(R)--+
|                                    |
|               +-----------+        |
|  status       |    TON    |        |
+----| |--------|EN      ENO|--      |
|               |           |        |
| NOT delay.Q --|IN        Q|--      |
|        T#1S --|PT       ET|--      |
|               +-----------+        |
|                                    |
|               +-----------+        |
|  delay.Q      |    ROL    |        |
+----| |--------|EN      ENO|--      |
|               |           |        |
|        LED8 --|IN      ERG|-- LED8 |
|           1 --|N          |        |
|               +-----------+        |
(End of Ladder Diagram)
```
\end{verbatim}

\subsubsection{Results}
The results are summarized in Table~\ref{tab:LD}.
\begin{table}[htbp]
    \caption{Scores for GPT4 generated LD}
    \centering
    \begin{tabular}{|c|c|c|c|c|}
    \hline
        \textbf{Number} & \textbf{CaseShortName} & \textbf{System Message} & \textbf{Step} & \textbf{Few-shot} \\ \hline
        \textbf{3\_1} & ConveyorControl & 0 & 0 & 0.5 \\ \hline
        \textbf{3\_2} & HeatingControl & 0 & 0 & 0 \\ \hline
        \textbf{3\_3} & TrafficControl & 0 & 0 & 0 \\ \hline
        \textbf{3\_4} & PneumaticControl & 0 & 0 & 0 \\ \hline
        \textbf{3\_5} & ElevatorControl & 0 & 0 & 0 \\ \hline
        \textbf{3\_6} & CarWash & 0 & 0 & 0.5 \\ \hline
        \textbf{3\_7} & CarPark & 0 & 0 & 0 \\ \hline
        \textbf{3\_8} & PickPlace & 0 & 0 & 0 \\ \hline
        \textbf{3\_9} & BottleRemoval & 0 & 0 & 0.5 \\ \hline
        \textbf{3\_10} & CoffeeMaker & 0 & 0 & 0 \\ \hline
    \end{tabular}
    \label{tab:LD}
\end{table}

\subsubsection{Analysis}
In the case of existing cases, no correct LD has been generated.
In few-shot learning, however, LD that is understandable by humans has been generated. To be more precise, it can be understood by engineers who have basic PLC programming knowledge.
Upon deeper analysis, we found that these cases are relatively simple cases.
Therefore, we have designed several more cases in subsequent sections (Experiment~3 and~4), simplified the logic, and provided more design details.

The results also demonstrate that GPT-4 struggles with understanding IEC 61131-3, because there are many errors in the symbols it provides. Many symbols do not exist in the standard, which means that GPT-4 has ``hallucinated'' these symbols.

One important reason for failure is that spaces are not counted as a token. We tested this on the tokenizer tool provided by OpenAI\footnote{\url{https://platform.openai.com/tokenizer}}. Adding or reducing spaces does not affect the number of tokens; it only affects the character count. However, spaces do affect ASCII art diagram, which impacts people's understanding of the LD.
Another reason may be that there is no specific fine-tuning specifically for ASCII art.

\subsubsection{Lessons Learned}
\begin{itemize}
    \item The performance difference between LD and ST for the same problem is like night and day. To simplify the requirements, we will try to propose more detailed specifications. Specifically, we'll abandon complexity, including elements such as timers, counters, etc., that may be difficult to understand for machine.
    \item Few-shot learning helps improve the results in some cases. For simple logic tests in Experiment~3 and~4, we will provide simple examples too.
\end{itemize}

\subsection{Experiment 2: PLC programs in Sequential Function Chart} \label{exp:SFC}
In this set of experiments, we repeat the experiments in Experiment~1. The difference is that the programming language will be changed to SFC.
In task decomposition, some requirements will be modified according to SFC language. Due to space limitations, they will not be displayed here.
In few-shot learning we also provide 3 examples, which are not demonstrated here for the same reason.

\subsubsection{Results}
The results are summarized in Table~\ref{tab:SFC}.
\begin{table}[htbp]
    \centering
    \caption{Scores for GPT4 generated SFC}
    \begin{tabular}{|c|c|c|c|c|}
    \hline
        \textbf{Case} & \textbf{CaseShortName} & \textbf{System Message} & \textbf{Step} & \textbf{Few-shot} \\ \hline
        \textbf{3\_1} & ConveyorControl & 0 & 0 & -1 \\ \hline
        \textbf{3\_2} & HeatingControl & 0 & 0 & 0.5 \\ \hline
        \textbf{3\_3} & TrafficControl & 0 & 0 & 0 \\ \hline
        \textbf{3\_4} & PneumaticControl & 0 & 0 & 0 \\ \hline
        \textbf{3\_5} & ElevatorControl & 0 & 0 & -1 \\ \hline
        \textbf{3\_6} & CarWash & 0 & 0 & 0.5 \\ \hline
        \textbf{3\_7} & CarPark & 0 & 0 & 0 \\ \hline
        \textbf{3\_8} & PickPlace & 0 & 0 & 0 \\ \hline
        \textbf{3\_9} & BottleRemoval & 0 & 0 & 0.5 \\ \hline
        \textbf{3\_10} & CoffeeMaker & 0 & 0 & 0 \\ \hline
    \end{tabular}
    \label{tab:SFC}
\end{table}

\subsubsection{Analysis}
The following analysis is based on few-shot learning.
Case 3\_1 and 3\_5 does not provide SFCs at all, they only describes SFCs in text, which does not comply with IEC 61131-3.
Case 3\_5 and 3\_8 try to identify synchronous paths, but failed.
Case 3\_6 and 3\_9 produced something resemble flowcharts rather than SFCs.

As for the reason, we believe it is similar to the previous experiment, namely that spaces are not counted as tokens.

\subsubsection{Lessons Learned}
\begin{itemize}
    \item For complex logic, GPT-4 cannot generate correct or useful SFCs.
    \item Few-shot learning helps improve the results.
    \item We need to find a way to treat spaces as valid tokens.
\end{itemize}

\subsection{Experiment 3: LD - Detailed Cases} \label{exp:LD_Detail}
In the previous experiments, whether using GPT-4 to generate LD or SFC, it can be considered as unsuccessful.
It seems that GPT is not competent enough for complex logic.
Therefore, we simplified the logical requirements of the design, removing potentially challenging logic such as timing and counting, and only examined the situation of discrete event logic.
There is also no parallel operation involved. Furthermore, we provided detailed names for each sensor and action logic.

Based on the conclusions from our previous experiments, decomposing tasks did not significantly increase accuracy, while few-shot learning significantly improved the outputs. Therefore, under these experimental conditions, we abandon task steps and adopt the combination of system messages with few-shot learning.

For this and subsequent experiments (i.e. Experiment~3 to~6), we designed 3 test cases that focus on discrete event systems with a limited number of states.
While in Case 1 the sequence is explained sequentially in the text (making it easier to interpret and convert to a discrete event based program), Case 3 has the sequence implicitly defined by the objectives that need to be achieved (making it more difficult to convert to a program). In terms of size complexity, case 1 involves two-bit logic while Case 2 involves only one-bit logic. Case 3 can be regarded as involving four-bit logic with 2 hidden states.

\begin{figure}[htbp]
    \centering
    \includegraphics[width=.8\linewidth]{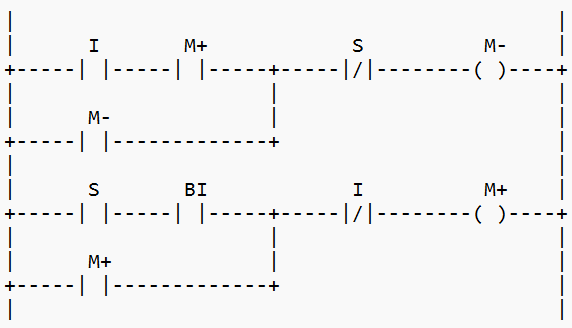}
    \caption{LD solution for Case 1}
    \label{fig:03_case1_ld}
\end{figure}

\begin{figure}[htbp]
    \centering
    \includegraphics[width=.8\linewidth]{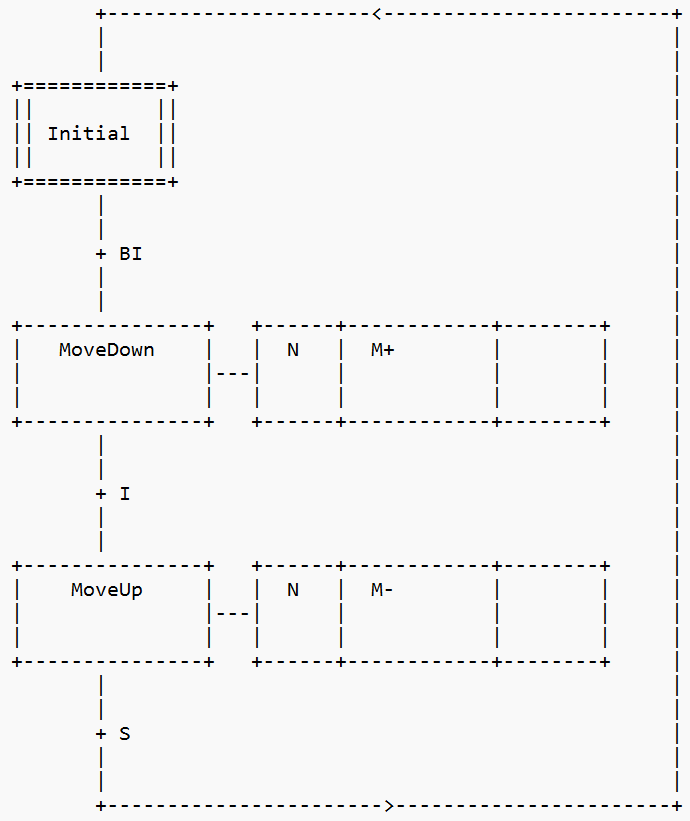}
    \caption{SFC solution for Case 1}
    \label{fig:03_case1_sfc}
\end{figure}

The prompts given to the LLM tool are provided below. For sake of comparison with LLM-generated code, we also provide examples of what we would consider a correct solution for Case 1: using LD in Figure~\ref{fig:03_case1_ld}, and SFC in Figure~\ref{fig:03_case1_sfc}.

\textbf{Case 1: Press Control} \label{case:01}
You want to control a hydraulic press. Consider the following specification:\\
- The system performs an operating cycle every time the start button (BI) is pressed.\\
- The press starts the cycle in the upper position (detected by sensor S going high), descending until it reaches the lower position (detected by sensor I going high).\\
- After the press reaches the lower position, the cycle continues by moving up until reaching the upper position, ending the cycle.\\
- The press is driven by a motor M with movement control in both directions: downward (M+) and upward (M-).

\textbf{Case 2: Motor Start / Stop} \label{case:02} Use start / stop button to control a motor. When the start button is pressed, the motor starts. When the stop button is pressed, the motor stops.

\textbf{Case 3: Two Pumps} \label{case:03} There are two water pumps (P1, P2) that pump out the water in a reservoir. To extend the lifespan of the water pumps, the two pumps operate alternately. When the water level is above the high water line (HL), the pump operates. When the water level is below the low water line (LL), the pump stops working.

\subsubsection{Results \& Analysis}
Due to the limited number of cases, we are not using a scoring mechanism. 
Actually, no correct LD solutions were generated - the severely mangled generated output attempting to mimic LD would take up considerable space, so we will not list the results here. For the complete results, please refer to the GitHub repository\footnote{\url{https://github.com/yimin-up/LLM_IEC-61131_Graphic}}.

The experimental results indicate that GPT-4 still cannot generate correct LD even for simple scenarios. This suggests that LD poses a significant challenge for LLMs.

\subsection{Experiment 4: SFC - Detailed Cases} \label{exp:SFC_Detail}
This experiment used the same cases and procedures as the previous Experiment~4, but now asking for results in SFC language.

\subsubsection{Results}
\begin{figure}[htbp]
    \centering
    \includegraphics[width=.8\linewidth]{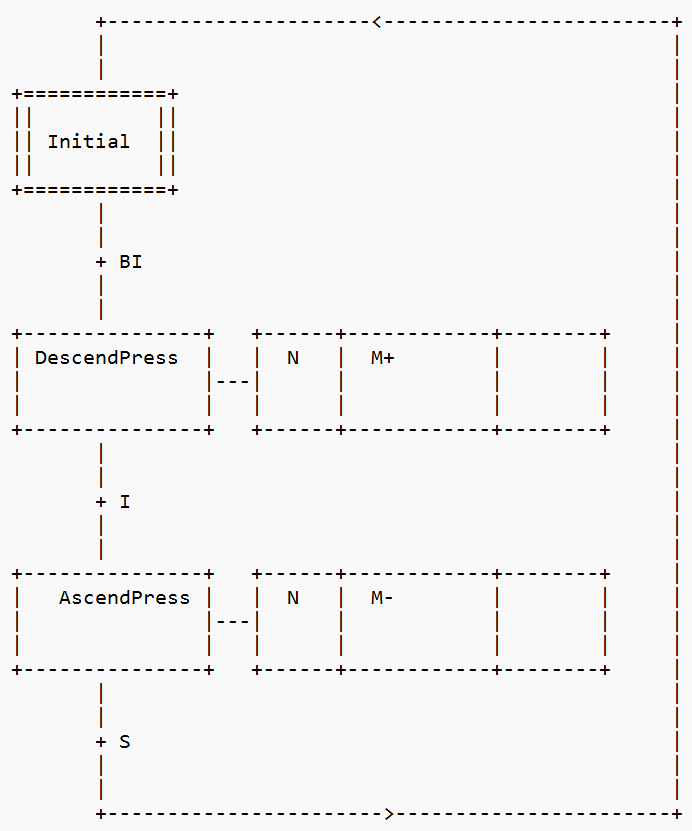}
    \caption{SFC solution generated by GPT-4 for Case 1}
    \label{fig:04_case1_sfc}
\end{figure}

\begin{figure}[htbp]
    \centering
    \includegraphics[width=.8\linewidth]{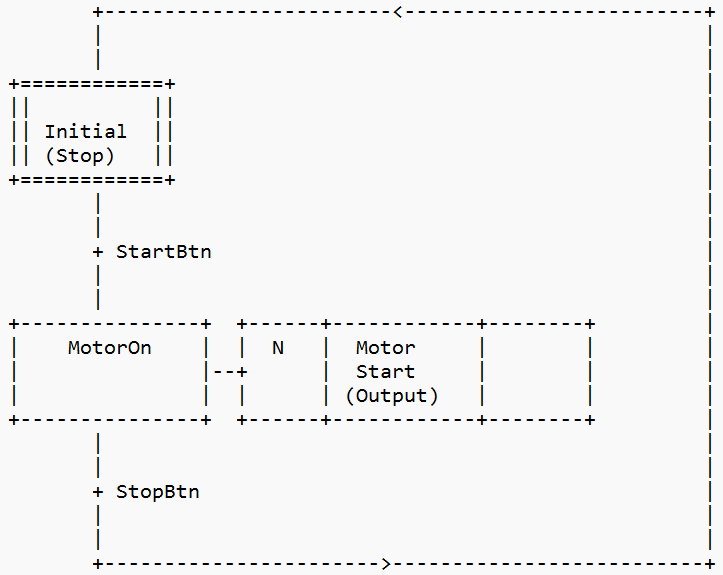}
    \caption{SFC solution generated by GPT-4 for Case 2}
    \label{fig:04_case2_sfc}
\end{figure}

\begin{figure}[htbp]
    \centering
    \includegraphics[width=.8\linewidth]{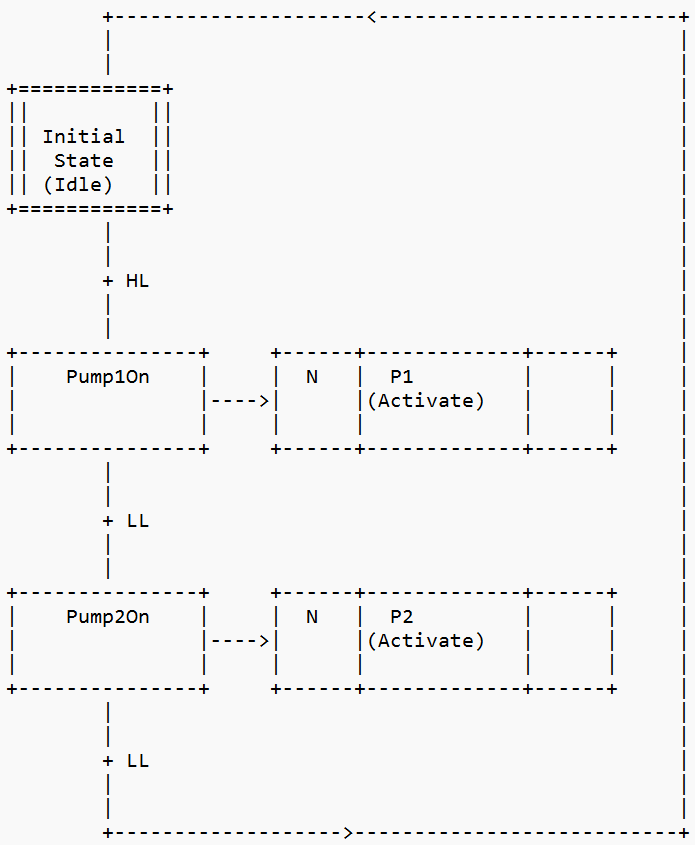}
    \caption{SFC solution generated by GPT-4 for Case 3}
    \label{fig:04_case3_sfc}
\end{figure}

\begin{figure}[htbp]
    \centering
    \includegraphics[width=.6\linewidth]{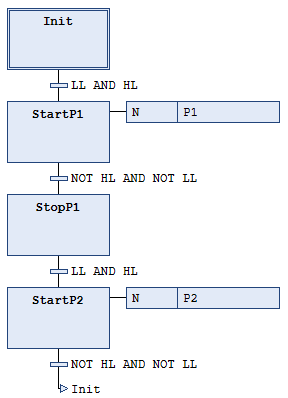}
    \caption{SFC solution in CODESYS for Case 3}
    \label{fig:04_case3_sfc_codesys}
\end{figure}

For Case 1 and 2, GPT-4 yielded completely semantically and syntactically correct results.
In Case 3 the hidden states were not generated, resulting in an incorrect outcome.
All results will be listed here (Figure~\ref{fig:04_case1_sfc}~-~Figure~\ref{fig:04_case3_sfc}).
The correct SFC for Case 3 is shown in Figure~\ref{fig:04_case3_sfc_codesys} for comparison.

\subsubsection{Analysis}
These outcomes are unexpected.
They indicate that SFC is more easily understood and learned by GPT-4.
However, if we repeat the same prompts several times, we get different outputs which may be incorrect.
Especially for Case 3, no correct answer was generated.
But for Case 1 and 2, containing 1-bit or 2-bit state machine, the output is correct.

\subsection{Experiment 5: LD-SFC Conversion} \label{exp:LD-SFC}
The conversion problem dates back to at least the 1990s when~\cite{Falcione1992} described it as a design recovery problem in 1992.
Since then, a few algorithms were proposed to convert LD into SFC, including graphical analysis~\cite{Falcione1992}, temporal logic methods~\cite{TadanaoTransformation}, state-space based algorithms~\cite{VimalUnderstanding}, etc.
In practice, these algorithms have not been widely adopted because they cannot solve issues such as a lack of domain knowledge and state space explosion.

In this experiment, we provide three conversion examples in the prompts, from LD to SFC, without any textual description of the problem so as to guarantee that the conversion doesn't come from the textual information.
The three examples are simple sequential sequences that loop back to the initial state, with 2, 3 and 4 states each. All examples provided use the same design pattern for implementing state machines using LD diagram, as exemplified in Figure~\ref{fig:03_case1_ld}.

We then provide the LD solution to each of the three test cases, and ask for the equivalent solution in SFC. It should be noted that the test cases are in essence identical to the examples given, with test case 1 having 3 sequential states, test case 2 having 2 states, and test case 3 having 4 sequential states. Once again we use the same design pattern for implementing state machines in LD diagram. These three test cases are however different to the examples given when taking into account the names of the variables (sensors and actuators). One of the examples also has some extra outputs (warning lights) that distinguishes it from the canonical 3-state sequential problem.

\subsubsection{Results \& Analysis}
No correct conversion was achieved.
The results will not be listed here. For more information, please refer to the GitHub repository. The results produced for the three examples resemble more of a flowchart, but even so incorrect in terms of identifying the number of states and their sequence; perhaps GPT-4 tends to generate flowcharts more often.

The difficulty in conversion may lie in three aspects:
\begin{itemize}
    \item Poor understanding of LD as Experiment~1 and~3 showed.
    \item The absence of auxiliary contextual information. As a language model, LLMs understand text better than ASCII art or graphical information.
    \item Lack corresponding corpora to training the model.
\end{itemize}
All of these aspects are worth further study.

\subsection{Experiment 6: SFC-LD Conversion} \label{exp:SFC-LD}
Contrary to the previous Experiment~5, there are rule-based solutions for converting a SFC into a LD. Theoretically, it's less challenging.
We once again provide the same three conversion examples in the prompts without any descriptions, and ask to convert the SFC solution to each of the test cases into equivalent LD programs.

\subsubsection{Results \& Analysis}
Unfortunately, no correct conversion was achieved.
Surprisingly, even for test case 2, which is a simple case with only one logical operation, resulted in a completely wrong output.
Although the graph generated for test case 3 is incorrect, it leads to the deduction of four states with 2 hidden state. It shows GPT-4 can understand the SFC in this case compared to Experiment~5. For this case, the correct LD could be obtained with slight manual modifications.
Due to space constrains, we refer the reader to the GitHub repository for the complete results.
As for possible reasons, we believe it's similar to the previous conclusion in Experiment~5.

\subsection{Experiment 7: GPT-4 vs Claude 3 vs Gemini} \label{exp:GPT_Claude_Gemini}
This experiment will compare results obtained when using the currently popular LLMs: GPT-4, Claude 3 and Gemini-1.5.
We adopt the few-shot learning strategy and use the scoring system mentioned earlier.
For ease of comparison, we tested on Category 3 (PLC Programming Exercises) using APIs.

\subsubsection{Results}
Table~\ref{tab:GPT_Claude_Gemini} shows the results from different LLMs. 
\begin{table}[!htbp]
    \centering
    \caption{Comparison between GPT-4, Claude 3 and Gemini-1.5}
    \begin{tabular}{|c|c|c|c|c|c|c|}
    \hline
        \multirow{2}*{\textbf{Case}} & \multicolumn{2}{|c|}{\textbf{GPT-4}} & \multicolumn{2}{|c|}{\textbf{Claude 3}} & \multicolumn{2}{|c|}{\textbf{Gemini-1.5}} \\ \cline{2-7}
        ~ & LD & SFC & LD & SFC & LD & SFC \\ \hline
        \textbf{3\_1} & 0.5 & -1 & 0 & -1 & 0 & 0 \\ \hline
        \textbf{3\_2} & 0 & 0.5 & -1 & -1 & 0.5 & 0.5 \\ \hline
        \textbf{3\_3} & 0 & 0 & -1 & -1 & 0 & 0 \\ \hline
        \textbf{3\_4} & 0 & 0 & 0 & -1 & 0 & 0 \\ \hline
        \textbf{3\_5} & 0 & -1 & -1 & 0 & 0 & 0 \\ \hline
        \textbf{3\_6} & 0.5 & 0.5 & 0.5 & 0.5 & 0 & 0.5 \\ \hline
        \textbf{3\_7} & 0 & 0 & -1 & 0.5 & 0 & 0 \\ \hline
        \textbf{3\_8} & 0 & 0 & -1 & 0 & 0 & 0 \\ \hline
        \textbf{3\_9} & 0.5 & 0.5 & 0 & 0 & 0 & 0.5 \\ \hline
        \textbf{3\_10} & 0 & 0 & -1 & 0 & 0 & 0.5 \\ \hline
    \end{tabular}
    \label{tab:GPT_Claude_Gemini}
\end{table}

\subsubsection{Analysis}
There is no clear winner or loser when comparing the results obtained, and all LLMs struggled to provide meaningful results. No LLM dominates the others, with each LLM able to provide a better results than the rest in at least one of the test cases. However it could be argued that Claude 3 is the weakest in this task because it more often doesn't provide a solution at all, or the solution provided is not in the required language.

\section{Conclusion and Future Work} \label{sec:conclusion}
We conducted a series of experiments to investigate the ability of LLMs in generating LD and SFC for PLC programming.
The results indicate that LLM is more capable of understanding SFC. However, for LD, LLM finds it relatively difficult to comprehend.
LLM can provide correct solutions for relatively simple state machines (1-bit, 2-bit) according to textual descriptions. However, for complex logic, it is currently incapable.
Moreover, regarding the conversion between LD and SFC, LLMs are not yet competent.

Future work includes Retrieval Augmented Generation (RAG) for LLMs, and fine-tune the models, trying to improving the accuracy of SFC, the correctness of LD, and the correctness in complex tasks.
Considering how to address spaces in ASCII art will be the next key focus.
Currently, all these experiments are black-box testings. Trying to find theoretical explanation for these results is another interesting direction.

\bibliographystyle{IEEEtran}


\end{document}